\documentstyle[12pt,epsfig]{article}
\newcommand{\beq}{\begin{equation}}
\newcommand{\eeq}{\end{equation}}
\newcommand{\bea}{\begin{eqnarray}}
\newcommand{\eea}{\end{eqnarray}}

\newcommand{\gsim}{\lower.7ex\hbox{$
\;\stackrel{\textstyle>}{\sim}\;$}}
\newcommand{\lsim}{\lower.7ex\hbox{$
\;\stackrel{\textstyle<}{\sim}\;$}}

\def\lsim{\mathrel{\rlap{\lower3pt\hbox{\hskip0pt$\sim$}}
    \raise1pt\hbox{$<$}}}         %less than or approx. symbol
\def\gsim{\mathrel{\rlap{\lower4pt\hbox{\hskip1pt$\sim$}}
    \raise1pt\hbox{$>$}}}         %greater than or approx. symbol

\renewcommand{\Im}{{\rm Im}\,}

\newcommand{\bibit}[1]{\bibitem{#1}}

\newcommand{\La}{\overline{\Lambda}}

\newcommand{\as}{\alpha_s}
\newcommand{\GeV}{\,\mbox{GeV}}
\newcommand{\MeV}{\,\mbox{MeV}}

\newcommand{\msp}[1]{\mbox{\hspace*{#1mm}~}}

\begin{document}

\begin{titlepage}
\renewcommand{\thefootnote}{\fnsymbol{footnote}}

\begin{flushright}
Bicocca-FT-02/22\\
hep-ph/0210413
\end{flushright}
\vspace*{1.3cm}

\begin{center} \Large

{\LARGE{\bf QCD \,Corrections \,in \boldmath $\;\Gamma_{\rm sl}(B)$}}
\end{center}
\vspace*{5mm}
\begin{center} {\Large\tt
Nikolai Uraltsev} \\
\vspace{1.2cm}
{\normalsize
{\it INFN, Sezione di Milano, Milan, Italy$^{\:*}$
%%% {\small \rm and} \\
%%% {\it ~\hspace*{-15mm} Department of Physics, University of Notre 
%%% Dame du Lac, Notre Dame, IN 46556, U.S.A.$^*$ \hspace*{-15mm}~}
}\\
}

\normalsize
\vspace*{35mm}

{\large{\bf Abstract}~~}\vspace*{-.1mm}\\
\end{center}
Short-distance expansion of the total semileptonic $B$ widths is
reviewed for the OPE-conformable scheme employing low-scale running quark 
masses. The third- and fourth-order BLM corrections are
given and the complete resummation of the BLM series presented. The
effect of higher perturbative orders with running quark masses is found very
small. Numerical consequences for $|V_{cb}|$ are addressed.
\vfill

~\hspace*{-7mm}\hrulefill \hspace*{3cm} \\
\footnotesize{\noindent $^*$On leave of absence from Department of 
Physics, University of Notre Dame, Notre Dame, IN 46556, U.S.A. and 
St.\,Petersburg Nuclear Physics Institute, Gatchina, St.\,Petersburg 
188300, Russia}

\noindent
\end{titlepage}

\newpage

%%% \section{Introduction}

Total decay rates of heavy flavors are dominated by short-distance physics;
nonperturbative effects originating from a typical hadronic momentum scale 
are described by the OPE. The semileptonic $B$ widths $\Gamma_{\rm
sl}(b\!\to\!c)$ and $\Gamma_{\rm sl}(b\!\to\!u)$ offer an unmatched
accuracy in determination of the corresponding CKM elements. To make a
full use of the infrared-free nature of perturbative corrections to
the widths they should be computed in the context of the Wilsonian
OPE; in particular this suggests using the running quark masses
normalized at a low scale $\mu \! \ll \! m_b$ \cite{pole}. The perturbative
corrections (i.e., the $\alpha_s$-expansion of the Wilson coefficients
of the unit operator $\bar{Q}Q$) explicitly depend in this approach on
the separation (normalization) scale $\mu$. 

The $\mu$-dependence of the Wilson coefficients is uniquely determined
\cite{optical,upset} by normalization point independence of the 
inclusive width and by known $\mu$-dependence of quark masses and 
local heavy
quark operators in the OPE series. Below the expressions are collected
for the scheme based on running `kinetic' heavy quark masses and other
operators defined via the small velocity (SV) sum rules
\cite{five,blmope}. These definitions are complete even beyond
perturbation theory; likewise they are valid to any perturbative
order.

\section{Perturbative corrections}

Considering pure short-distance (perturbative) effects, we denote
\bea
\nonumber
\Gamma_{\rm sl}^{\rm pert}(b\!\to\!c)\msp{-3} &=& \msp{-3}\frac{G_{F\,}^2
m_b^5(\mu)}{192\,\pi^3}\; \raisebox{-.5mm}{\mbox{{\large$|V_{cb}|^2$}}} 
\,\left(1\!+\!A_{\rm ew}\right)\,
\left\{z_0(r) \,+ 
\mbox{$\,\bar a_1(r;\mu)\frac{\alpha_s(m_b)}{\pi} \,+\,  \bar a_2(r;\mu)
\left(\frac{\alpha_s(m_b)}{\pi}\right)^2$} \right.\;\;\\
&& \msp{-3} \left. \mbox{$ + \: \bar a_3(r;\mu)
\left(\frac{\alpha_s(m_b)}{\pi}\right)^3 + ...$}
\right\}\;,
\qquad \qquad \label{108}
\eea
where $z_0(r)$ is the tree-level phase space factor and 
$r\!=\!m_c^2(\mu)/m_b^2(\mu)$:
\beq
z_0(r)=1-8r+8r^3-r^4-12r^2\ln{r}\;.
\label{110}
\eeq
In what follows $\alpha_s$ is the $\overline{{\rm MS}}$ strong
coupling normalized at $m_b$, unless indicated explicitly. 
 
The electroweak corrections $A_{\rm ew}$ describe the ultraviolet
renormalization of the semileptonic Fermi interaction and, being the 
QED-counterpart of the familiar factors $c_\pm$ in nonleptonic Lagrangian,
has a well-known logarithm of $M_Z/m_b$ due to virtual photon
exchanges. It depends only on the fermion charges and is the same as in
$\beta$-decays of neutron or hyperons \cite{ewlog}:
\beq
1+A_{\rm ew}\simeq \left(1+\frac{\alpha}{\pi} \,\ln{\frac{M_Z}{m_b}}\right)^2.
\label{106}
\eeq

The $\mu$-dependence of the perturbative factors $\bar a_1(r;\mu)$ and
$\bar a_2(r;\mu)$ through the second order in $\alpha_s$ is given by 
\bea
\bar a_1(r;\mu)\msp{-3} &=& \msp{-3} 
\bar a_1^{(0)}(r) + \left[5\Lambda_1+3p_1 \right]\,z_0(r)+ 
\left[2(\sqrt{r}\!-\!r)\Lambda_1+(1\!-\!r)p_1 \right]\,
\frac{{\rm d}z_0(r)}{{\rm d}r}\label{127}
\\
\nonumber
\bar a_2(r;\mu)\msp{-3} &=& \msp{-3} 
\bar a_2^{(0)}(r) + \left[5\Lambda_1+3p_1 \right]\,a_1^{(0)}(r)
+\left[2(\sqrt{r}\!-\!r)\Lambda_1+(1\!-\!r)p_1 \right]\,
\frac{{\rm d}a_1^{(0)}(r)}{{\rm d}r}
\\
\nonumber
&& \msp{-12}
+ \left[5\Lambda_2+3p_2 +10\Lambda_1^2+4p_1^2 +
\frac{25}{2}\Lambda_1 p_1\right]\,z_0(r) \\
\nonumber
&& \msp{-12}
+\left[2(\sqrt{r}\!-\!r)\Lambda_2+(1\!-\!r)p_2+
(1\!+\!6\sqrt{r}\!-\!7r)\Lambda_1^2 +(\frac{1}{4r}\!+\!2\!-\!\frac{9}{4}r)p_1^2
\:+ \right. \\
%%% \nonumber
&& \msp{-6} %% \left.
(\frac{1}{\sqrt{r}}\!+\left.\!3\!+\!4\sqrt{r}\!-\!8r)\Lambda_1 p_1
\rule{0mm}{5mm} \right]\,\frac{{\rm d}z_0(r)}{{\rm d}r} 
%%% \\ && \msp{-12} 
+ \frac{1}{2}
\left[2(\sqrt{r}\!-\!r)\Lambda_1+(1\!-\!r)p_1 \right]^2
\frac{{\rm d}^2z_0(r)}{{\rm d}r^2}\,,\qquad
\label{128}
\eea
where $\bar a_k^{(0)}(r)\!\equiv \!\bar a_k(r;0)$ are the
coefficients perturbatively computed at $\mu\!=\!0$, i.e.\ for the
pole quark masses and the pole-type definition of the kinetic and
higher-order operators. 
The $b\!\to\! u$ case corresponding to $r\!=\!0$ amounts to using
$z_0\!=\!1$, $a_1\!=\!-\frac{2}{3} \left(\pi^2\!-\!\frac{25}{4}\right)$ while 
discarding all terms with derivatives.

Four parameters $\Lambda_{1,2}$ and $p_{1,2}$ entering above denote the
coefficients in the perturbative expansion of $\La(\mu)/m_b$ and 
$\mu_\pi^2(\mu)/m_b^2$ to first and second order in
$\alpha_s(M)/\pi$, respectively. In the
the kinetic mass scheme they are \cite{dipole}
\bea
\label{111}
\Lambda_1\msp{-3} &=& \msp{-3}\frac{4}{3}C_F\frac{\mu}{m_b}, \qquad
\Lambda_2=\Lambda_1\left[\frac{\beta_0}{2}\left(\ln{\frac{M}{2\mu}}+\,
\frac{8}{3}\,\right) 
- C_A\left(\frac{\pi^2}{6}\!-\!\frac{13}{12}\right)\right]\,, \\
p_1\msp{-3} &=& \msp{-3} \;\;C_F\frac{\mu^2}{m_b^2}, \qquad \;
p_2=\,p_1 \left[\frac{\beta_0}{2}\left(\ln{\frac{M}{2\mu}}+
\frac{13}{6}\right) 
- C_A\left(\frac{\pi^2}{6}\!-\!\frac{13}{12}\right)\right]. \qquad \qquad
\label{112}
\eea
Here 
$$
C_F=\frac{4}{3}\,, \qquad C_A\!=\!N_c\!=\!3\,, \qquad
\beta_0=\frac{11}{3}C_A-\frac{2}{3} n_f=9\;.
$$

The general structure of Eqs.~(\ref{127}) and (\ref{128}) 
holds in higher orders in $\alpha_s$,  
but the explicit 
coefficients change and the terms proliferate. This does not
happen for the BLM corrections\,\footnote{Perturbative subseries
including the coefficients of the form
$a_n=\left(\frac{\beta_0}{2}\right)^{n\!-\!1}\tilde a_n$, where the
coefficients are viewed as polynomials in $n_f$.} to any 
order, for which the translation between different $\mu$ takes 
basically the same form as for
the first-order coefficient $a_1(r;\mu)$. The only dependence on the
order $k$ in the BLM expansion is the constant accompanying the power
of the renorm-group $\log$s 
$\left(\ln{\frac{m_b}{2\mu}\!+\!\frac{8}{3}}\right)^{k\!-\!1}\!\!\!\!\;$, 
$\,\;\left(\ln{\frac{m_b}{2\mu}\!+\!\frac{13}{6}}\right)^{k\!-\!1}\;$ in 
Eqs.~(\ref{111}), (\ref{112}) 
(details can be found in Ref.~\cite{blmope}).
For the third order BLM, in particular, they were given in
\cite{dipole}:
\beq
\Lambda_3^{\mbox{\tiny BLM}}=\Lambda_1\,
\mbox{$\left(\frac{\beta_0}{2}\right)^2\!
\left[\left(\ln{\frac{M}{2\mu}}\!+\! \frac{8}{3}\right)^2 
\!+\frac{67}{36}\!-\!\frac{\pi^2}{6}\right]$}\,, \;\;\;\; 
p_3^{\mbox{\tiny BLM}}=p_1\,
\mbox{$\left(\frac{\beta_0}{2}\right)^2\!
\left[\left(\ln{\frac{M}{2\mu}} \!+\! \frac{13}{6}\right)^2
\!+\frac{10}{9}\!-\!\frac{\pi^2}{6}\right]$}\,.
\label{132}
\eeq
Although the expressions for even 
higher orders in BLM is straightforward, in practice it is simpler 
to evaluate the BLM
corrections in the widths directly in the running mass scheme. 
The general formalism is described in \cite{blmope}, and using
an additional technical trick given in Appendix  
simplifies necessary integrations.

\section{Numerical estimates}

For the second-order {\tt non-BLM} coefficient we have 
\beq
\bar a_2(r;\mu) -\bar a_2^{(0)}(r) 
\raisebox{-.9mm}{\mbox{$\left|\rule{0mm}{3.5mm} \right.$}} 
\raisebox{-2mm}{\mbox{$_{\rm non-BLM}$}} 
\equiv \delta_2^{(0)}(r;\mu)\;, \qquad
\delta_2^{(0)}(0.25^2;1/4.6)\simeq -1.35
\label{140}
\eeq
The simple interpolating expression accounting for the mass dependence
of $\bar a_1^{(0)}$, 
\bea
\nonumber
\delta_2^{(0)}(r;\mu)\msp{-3}& \simeq& \msp{-3} 
-\frac{9.9\:\mu}{m_b}\,z_0(r)\,\left[1
\!-\!4.5\,(\sqrt{r}\!-\!0.25)+13\,(\sqrt{r}\!-\!0.25)^2 
 +0.7\,\left(\frac{\mu}{m_b}\!-\!\frac{1}{4.6}\right)\right.\\
& & \msp{-3} \left.
-12\,(\sqrt{r}\!-\!0.25)\left(\frac{\mu}{m_b}\!-\!\frac{1}{4.6}\right)+
71\,(\sqrt{r}\!-\!0.25)^2\,\left(\frac{\mu}{m_b}\!-\!\frac{1}{4.6}\right)
  \right]
\qquad \label{142}
\eea
is accurate enough in the relevant domain of quark masses and
factorization scale $\mu$.\footnote{Recent analysis by DELPHI
\cite{delphi} yielded $m_b(1\GeV)\!\simeq\! 4.6\GeV$ and 
$r\!\approx\!0.06$.} 

Using the evaluation $\bar a_2^{(0)}\!\simeq\! 0.9$ of the second-order
non-BLM coefficients by Czarnecki and
Melnikov \cite{czarmint} we obtain
\beq
\bar a_2^{\rm non-BLM}(0.25^2;1\GeV)\simeq -0.65 \, z_0(0.25^2)
\label{143}
\eeq

The second-order BLM coefficient has been addressed in the literature
more than once;
here the shift between $\mu\!=\!0$ and $\mu\!=\!1\GeV$
constitutes about $\,0.78\,\beta_0\,$ for 
$r\!\simeq\!0.25^2$, and  
\beq
\bar a_2^{{\rm BLM}}(r;\mu)\simeq -0.69\,\beta_0\:z_0(r)\;\;
\mbox{ at } r=0.25^2 \mbox{ and } \frac{\mu}{m_b}=\frac{1}{4.6}
\label{144}
\eeq

All BLM coefficients can be readily computed
numerically along the lines of Ref.~\cite{blmope} using 
Eq.~(\ref{424}) given below in Appendix.
The required expression for the first-order perturbative
correction with non-zero gluon mass  is given in 
Ref.~\cite{bbbsl}.
This yields
\beq
\bar a_3^{{\rm BLM}}(0.25^2;\mu)\!\simeq\! -0.52\,\beta_0^2\,z_0(r)\,,
\;\;\;\; 
\bar a_4^{{\rm BLM}}(0.25^2;\mu)\!\simeq\!  -0.27\,\beta_0^3\,z_0(r)\;\;\;
\mbox{ at } \frac{\mu}{m_b}\!=\!\frac{1}{4.6}\;
\label{146}
\eeq

BLM-wise, it is not difficult to perform the all-order resummation of
the perturbative corrections in terms of the running masses. The
result reads
\bea
\nonumber
A^{\rm BLM}(r;\mu) \msp{-3} & \equiv &\msp{-3} \frac{1}{z_0(r)}
\left\{z_0(r) + \mbox{$\bar a_1(r;\mu)\frac{\alpha_s(m_b)}{\pi} +  \bar
a_2^{\rm BLM}(r;\mu)\left(\frac{\alpha_s(m_b)}{\pi}\right)^2$} \right.\\
&& \msp{40}
\left. \mbox{$ +  \bar a_3^{\rm BLM}(r;\mu)
\left(\frac{\alpha_s(m_b)}{\pi}\right)^3 + ...$}
\right\} \simeq 0.880\,;
\qquad \label{160}
\eea
it is independent of the scale for $\alpha_s$ one starts
with. 

The distribution of the gluon virtualities $\,Q\,$ in $\,\Gamma_{\rm
sl}(b\!\to\! c)\,$ is illustrated in Fig.~1 in the case of pole masses
($\mu\!=\!0$) and for $\mu\!=\!1\GeV$, $\mu\!=\!1.5\GeV$ and
$\mu\!=\!2\GeV$. The effect of removing the infrared domain is self-manifest.
Moreover, it is evident that a too significant part of the perturbative
corrections with pole masses comes from gluon momenta below $500\MeV$,
which would then cast legitimate doubts on the numerical results.

\begin{figure}[hhh]
\begin{center}
\mbox{\psfig{file=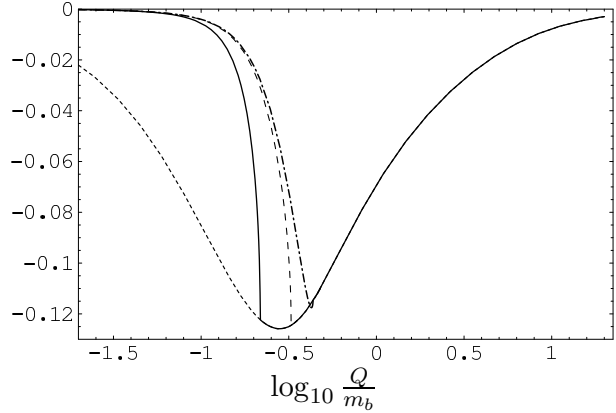,width=8cm}}
\end{center}\vspace*{-5mm}
\caption{{\small 
Gluon momentum scale distribution in 
$\Gamma_{\rm sl}(b\!\to\!c)$. Solid, dashed and dot-dashed lines
correspond to $\mu\!=\!1\GeV$, $\,1.5\GeV\,$ and $\,2\GeV$,
respectively; lighter short-dashed line illustrates the case of 
$\mu\!=\!0$ (pole masses). The area under each curve gives 
the first-order perturbative coefficient.}}
\end{figure}

The growing component of the BLM series becomes sign-alternating
starting the third or fourth order, this is a result of the infrared-free
form of the width in the OPE-motivated approach. Consequently, the
second-order approximation already turns out quite accurate, and the
final number is practically reproduced by including the
third-order term. It is also worth noting that using instead of 
$\alpha_s^{\overline{\mbox{\tiny MS}}}(m_b)$ a better physically
justified expansion parameter \cite{look,upset}, for example the
$V$-scheme $\alpha_s^{(V)}(m_b)$ or the `dipole radiation' 
$\alpha_s^{(\omega)}(m_b)$ yields a nearly precise perturbative factor
already to the first order:
\beq
A^{(V)}(r;\mu)\simeq (1-0.105-0.018+0.017-...)\;.
\label{164}
\eeq
\vspace*{2mm}

The case of $\Gamma_{\rm sl}^{\rm pert}(b\!\to\!u)$ is technically
simpler. Here the full second-order corrections are known analytically
\cite{timo}, $\;a_2^{(0)}(0)\!\simeq\! 5.54$, \,we have 
$\;\delta_2^{(0)}\!\simeq\! -6.9\:$ and, therefore
\beq
\bar a_1(0; 1\GeV)\simeq -0.29 \,, \;\;\;
\bar a_2^{\rm non-BLM}(0; 1\GeV)\simeq -1.35 \,,\;\;\; 
\bar a_2^{\rm BLM}(0; 1\GeV)\simeq 0.44\,\beta_0\,;
\label{172}
\eeq
third- and fourth-order BLM corrections amount to
\beq
\bar a_3^{\rm BLM}(0; 1\GeV)\simeq 1.23\,\beta_0^2 \,, \qquad
\bar a_4^{\rm BLM}(0; 1\GeV)\simeq 2.2\,\beta_0^3\:.
\label{174}
\eeq
A meaningful BLM summation for $b\!\to\!u$ would require, however
incorporating the four-fermion operator \cite{WA} 
$\;\bar{b}\gamma_\mu(1\!-\!\gamma_5)u\,\bar{u}\gamma_\nu(1\!-\!\gamma_5)b\;
(\delta_{\mu\nu}\!-\!\delta_{\mu 0}\delta_{\nu 0})\,$ as well -- 
it is strongly enhanced and has significant negative
anomalous dimension. Accounting for this operator affects already the tree
phase space coefficient at order $\mu^3/m_b^3$.

\section{\boldmath $\Gamma_{\rm sl}(B)$ }

In the complete OPE prediction for the semileptonic width,
there remains a minor arbitrariness in incorporating the power
corrections from the chromomagnetic and Darwin terms until their
coefficients are computed with the ${\cal O}(\alpha_s)$ 
accuracy.\footnote{The
coefficient for the kinetic operator equals to the one in the parton term,
while for the $LS$ operator it is a combination of the latter and the
coefficient for the chromomagnetic operator, see
Ref.~\cite{optical}.} There are reasons to think that 
the factorized form has some advantages:
\beq
\Gamma_{\rm sl}(B)=\Gamma_0 A^{\rm pert}\!\left[z_0(r)\!\left(
1\!-\!\frac{\mu_\pi^2\!-\!\mu_G^2 \!+\! 
\frac{\tilde\rho_D^3\!+\!\rho_{LS}^3}{m_b}}{2m_b^2}\right)
-2(1\!-\!r)^4 
\frac{\mu_G^2\!-\!\frac{\tilde\rho_D^3\!+\!\rho_{LS}^3}{m_b}}{m_b^2}+
D(r)\,\frac{\tilde \rho_D^3}{m_b^3}+ \!...
\right]\!,
\label{300}
\eeq
where $D(r)$ can be obtained from Ref.~\cite{grekap}:
\beq
D(r) =  8\ln{r} + \frac{34}{3}-\frac{32}{3}r -8 r^2
+\frac{32}{3}r^3 -\frac{10}{3} r^4\;\simeq\; -18.3 \,z_0
\qquad \label{32}
\eeq
The kinetic and chromomagnetic expectation values here are
finite-$m_Q$ matrix elements in actual $B$ mesons. (As argued in
Ref.~\cite{chrom}, their shift compared to the limit
$m_Q\!\to\!\infty$ is strongly suppressed.)

To give the numerical prediction, we evaluate this at the
canonical values $m_b(1\GeV)\!=\!4.6\GeV$,
$\mu_\pi^2(1\GeV)\!=\!0.4\GeV^2$, $\mu_G^2(1\GeV)\!=\!0.35\GeV^2$,  
$\tilde\rho_D^3(1\GeV)\!=\!0.12\GeV^3\,$ and 
$\,\rho_{LS}^3(1\GeV)\!=\!-0.15\GeV^3\:$ using $\:r\!=\!0.25^2\,$ and
$\,\alpha_s(m_b)\!=\!0.22\,$:
\beq
|V_{cb}|\simeq V_{cb}^{(0)}\; \mbox{{\large$
\left(\frac{{\rm Br}_{\rm sl}(B)}{0.105}\right)^{\!\frac{1}{2}} 
\left(\frac{1.55{\rm ps}}{\tau_B}\right)^{\frac{1}{2}}$}}
\left(1\!-\!\mbox{{\small$4.8\,[{\rm Br}(B\!\to\!X_u\,\ell\nu)-0.0018]$}}
\right) 
, \msp{4}
V_{cb}^{(0)}\simeq 0.0421\,.
\label{342}
\eeq
The above evaluation uses the full second-order perturbative
corrections with BLM resummation Eq.~(\ref{160}), where the added 
non-BLM term is computed at $\alpha_s\!=\!0.25$. The resulting value
of $V_{cb}^{(0)}$ acquires extra factors $\,0.993\,$, $\,1.002\,$  or
$\,1.004\,$ if one discards $\,{\cal O}(\alpha_s^3)\,$ and higher terms, 
including only the third-order BLM correction, or both third and 
fourth orders, respectively.
The dependence on the heavy quark parameters is as follows \cite{amst}:
{\small
\begin{eqnarray}
|V_{cb}| \msp{-3}&=&\msp{-3} V_{cb}^{(0)}\times
[1+0.4\,(\alpha_s(m_b)\!-\!0.22)]\, \times \nonumber 
\\ 
& & \msp{22}
\left[1 -0.65 \left ( m_b(1\GeV) \!-\!4.6\GeV \right )
+0.40 \left ( m_c(1\GeV) \!-\!1.15\GeV \right )\qquad \right. \nonumber \\
& & \msp{29}+0.01 \left ( \mu_{\pi}^2 \!-\!0.4\GeV^2 \right )
+0.10 \left ( \tilde\rho_D^3 \!-\!0.12\GeV^3 \right ) \nonumber \\
& & \msp{29}\left. +\,0.05 \left ( \mu_G^2\!-\!0.35\GeV^2 \right ) 
-0.01 \left ( \rho_{LS}^3 \!+\!0.15\GeV^3 \right ) \right ].
\label{346}
\end{eqnarray}
}

It should be noted that the Darwin expectation value
$\,\tilde\rho_D^3\,$ appearing above is not a true matrix element
normalized at $1\GeV$, but is rather understood as the one
extrapolated to $\,\mu\!=\!0\,$: to two loops $\,\tilde\rho_D^3 \!\approx\!
\rho_D^3(1\GeV)\!-\!0.1\GeV^3$; the perturbative
coefficients have been determined correspondingly. 
Modifying equations for using $\,\rho_D^3(1\GeV)\,$  proper is
straightforward. This does not change the
value of $\,|V_{cb}|\,$ by any appreciable amount. The only drawback of
the adopted simplified option is that
the apparent extracted value of $\,\tilde\rho_D^3\,$ is strongly
correlated with the approximation applied.\vspace*{3mm}

{\it To summarize,} \,the perturbative corrections to the total
semileptonic widths of beauty particles are modest and
allow precision theoretical control when viewed as the Wilson
coefficient of the leading heavy quark operator in the OPE. This
assumes using well-defined low-scale running masses; these masses can be
accurately extracted from experiment. As was pointed out long ago
\cite{upset}, the actual short-distance perturbative corrections to
semileptonic widths are only about 10 percent, and can be accurately
evaluated already at the one-loop  
level adopting a
physically motivated perturbative coupling. 
\vspace*{2.5mm}

{\bf Acknowledgments:} \,\,I am grateful to  
M.\,Battaglia, D.\,Benson, I.\,Bigi, P.\,Gam\-bino, T.\,Mannel 
and P.\,Roudeau for 
useful discus\-sions. 
This work was supported in part by the NSF under grant number PHY-0087419.
\vspace*{10mm}

\noindent
{\Large\bf Appendix} \vspace*{.5cm}
\renewcommand{\theequation}{A.\arabic{equation}}
\setcounter{equation}{0}

Here a brief description of technicalities involved is given following
Ref.~\cite{blmope}. Computing the
BLM corrections is most straightforward in the generalized dispersive
approach \cite{dmw} employing the dispersion relation for the dressed
gluon propagator:
\beq
\frac{\as(k^2)}{k^2} \left(\delta_{\mu\nu}-c\frac{k_\mu k_\nu}{k^2}\right)\;=\;
-\frac{1}{\pi}\;\Im \int \;\frac{d\lambda^2}{\lambda^2}\;\frac{\Im
\as(-\lambda^2)}{k^2+\lambda^2}\; \left(\delta_{\mu\nu}-c\frac{k_\mu
k_\nu}{k^2}\right)\;.
\label{b6}
\eeq
The form of the integrand suggests considering for a generic
observable $A$ the one-loop diagram with a non-zero gluon mass $\lambda$ 
\beq
A(\lambda^2)=1+\frac{\as}{\pi}A_1(\lambda^2)
\label{b22}
\eeq
and integrating it over $\lambda^2$ with the weight
$\rho(\lambda^2)\frac{{\rm d}\lambda^2}{\lambda^2}= -\frac{\!1}{\pi^{2\!}}
\,\Im \as(-\lambda^2)\;\frac{{\rm d}\lambda^2}{\lambda^2}$:
\beq
A^{\rm BLM}\;=\;1+ 
\int \;\frac{{\rm d}\lambda^2}{\lambda^2}\;A_1(\lambda^2)\, 
\rho(\lambda^2)\;\;.
\label{b23}
\eeq

A simple trick appears useful. Expressing 
$$A_1(\lambda^2)=
A_1(0)\frac{M^2}{M^2+\lambda^2} - 
\left(A_1(0)\frac{M^2\!}{M^2\!+\!\lambda^2} \!-\! A_1(\lambda^2)
\right)$$ 
with an arbitrary $M$, the
integral of the first term yields identically
$A_1(0)\frac{\alpha_s(M^2)}{\pi}$ and, therefore
\beq
A^{\rm BLM}\;=\;1+ A_1(0)\frac{\alpha_s(M^2)}{\pi}-
\int \;\frac{{\rm d}\lambda^2}{\lambda^2}\, \rho(\lambda^2)
\;\left(A_1(0)\frac{M^2\!}{M^2\!+\!\lambda^2} \!-\! A_1(\lambda^2)\right)
\;.
\label{23a}
\eeq

The standard BLM expansion amounts to using literal one-loop
$\alpha_s(k^2)$:
\beq
\alpha_s(k^2)=\frac{\alpha_s(Q^2)}{1+ \frac{\beta_0\alpha_s(Q^2)}{4\pi} 
\ln{\frac{k^2}{Q^2}}}\;.
\label{211}
\eeq
As had been shown in \cite{bbb6}, in this setting the dispersive approach
reproduces the literal BLM series. 

Taking  $\,M^2\!=\!e^{\frac{5}{3}}\,m_Q^2\;$  we obtain expansion directly
in terms of
$\alpha_s^{\overline{\mbox{\tiny MS}}}(m_Q)\,$: 
\bea
\nonumber
A^{\rm BLM}\msp{-3}& =& \msp{-3} 1+
A_1(0)\frac{\alpha_s(m_Q)}{\pi}\\
\nonumber
& & \msp{-.7}+
\int_{-\infty}^{\infty} \; {\rm d}t\:  
\frac{\frac{\beta_0}{4}\,\left(\frac{\alpha_s}{\pi}\right)^2}{\left(1+ 
\frac{\beta_0\alpha_s}{4\pi}(t\!-\!\frac{5}{3})\right)^2+
\left(\frac{\beta_0}{4}\alpha_s\right)^2} 
\:\left( A_1(0)\frac{1}{1+e^{t-\frac{5}{3}}}\!-\! A_1(e^{t}m_Q^2)\right)
\\
& & \msp{-.7}
- \frac{4}{\beta_0}\left[\frac{m_Q^2}{m_Q^2\!-\!\Lambda_V^2}A_1(0)
-A_1(-\Lambda_V^2)
\right]
\;,
\label{420}
\eea
where $\alpha_s\equiv\alpha_s^{\overline{\mbox{\tiny MS}}}(m_Q)$,
and 
\beq
\Lambda_V^2\,=\,m_Q^2 \:e^{-\frac{4\pi}{\beta_0\alpha_s(m_Q)}+\frac{5}{3}}\;.
\label{422}
\eeq
This form is convenient for numerical integration. It also allows the
straightforward expansion in $\alpha_s(m_Q)$:
\bea
\nonumber
A^{\rm BLM}\msp{-3}& =& \msp{-3}1\;+\; A_1(0)\frac{\alpha_s(m_Q)}{\pi}
\;+\;
\sum_{n=0}^{\infty}\: 
\mbox{$\frac{4}{\beta_0}
\left(\frac{\beta_0\alpha_s(m_Q)}{4\pi}\right)^{n+2}$} \times
\;\\
& & \msp{-3}
\sum_{k=0}^{\frac{n}{2}}\, 
(-\pi^2)^k \;\mbox{\large$\raisebox{-.5mm}{$C$}_{_{\,n+1}}^{^{\,2k+1}}$}
\cdot\int
\!\frac{{\rm d}\lambda^2}{\lambda^2}\:\left[\ln{\frac{m_Q^2}{\lambda^2}}
\!+\!\frac{5}{3}\right]^{n-2k}
\!\mbox{$\left(A_1(0)\mbox{\large$\frac{m_Q^2}{m_Q^2+ 
e^{\mbox{\tiny -5/3}}\lambda^2}$}\!-\! 
A_1(\lambda^2) \right)$}\,
.\qquad\;\;\;\;
\label{424}
\eea
Since the infrared part is removed from the one-loop diagram for the
total width, the corresponding $A_1(\lambda^2)$ is a real analytic
function in the vicinity of zero, and no ambiguity in Eq.~(\ref{420})
is encountered. Nevertheless, as follows from the generating integral
(\ref{420}) the series (\ref{424}) is only asymptotic.

Computing BLM corrections with the running masses $m_Q(\mu)$ requires 
expressions for the order-$\alpha_s$ terms in $\La^{\:\mbox{\tiny pert}}(\mu)$
and $\mu_\pi^{2\;\mbox{\tiny pert}}(\mu)\,$. They are given by
integrating the SV spectral density (Eqs.~(16), (19) of
Ref.~\cite{blmope}) with the proper power of energy:
\bea
\nonumber
\La^{\:\mbox{\tiny pert}} (\mu;\,\lambda)\msp{-3}&=&\msp{-3} 
\frac{16\alpha_s}{9\pi}\;
\vartheta \,(\mu^2\!-\!\lambda^2)\,\left[
(1\!-\!\mbox{$\frac{\lambda^2}{4\mu^2}$})\sqrt{\mu^2\!-\!\lambda^2}
 \!-\!\frac{3\pi}{8}\lambda+ 
\frac{3}{4}\lambda 
\arctan\frac{\lambda}{\sqrt{\mu^2\!-\!\lambda^2}}
\right]\,,\\
\mu_\pi^{2\;\mbox{\tiny pert}} (\mu;\,\lambda) \msp{-3}&=&\msp{-3} 
\;\frac{4\alpha_s}{3\pi}\;\,
\vartheta \,(\mu^2\!-\!\lambda^2) \;
\frac{(\mu^2\!-\!\lambda^2)^{3/2}}{\mu}\;\;.
\label{428}
\eea

As discussed in Refs.~\cite{bbb6,dmw}, the quantity 
$-\lambda^2\frac{{\rm d}A_1(\lambda^2)}{\lambda^2}$ can be 
qualitatively viewed as describing the
contribution of gluons with virtuality $\lambda^2$ in the one-loop
process. This convention has been adopted in Fig.~1.

\end{document}